# Ten questions and answers about superconductivity

T. D. Cao

*Department of Physics,*
*Nanjing University of Information Science & Technology, Nanjing 210044, China*

**This work answers the basic questions of superconductivity in a question-and-answer format. We extend a basic hypothesis to various superconductors. This hypothesis is that superconductivity requires that the pairing gap locates around the Fermi level. On the basis of this hypothesis our calculations give the so-called three factor theory with which some key problems of the high temperature superconductivity are explained.**

1. How to judge whether superconductivity is included in a model?

To answer this question, we should know the conditions appearing superconductivity, while this requires a hypothesis. A basic hypothesis is that superconductivity requires that the pairing gap should appear around the Fermi level. It is necessary to note that pairing functions are also limited around the Fermi level if the pairing gap is limited around the Fermi level, therefore, superconductivity is due to the pairing around the Fermi level. This hypothesis is not a pure guess because it can be found in the well-known BCS equation [1] and other similar equations, and we just extend this Cooper's idea [2] to various superconductors. The Fermi level is defined as the positions of the chemical potential in the wave vector space at a temperature. This means that the answer can be found by calculating the correlation functions in the wave vector space.

One often judge whether superconductivity is included in a model in the real space, but whether the pairing gap exists around the Fermi level could not be determined by the off-diagonal long-range order which was presented by Yang [3]. In addition, the superconductivity is not due to the BEC of bosons. These conclusions have not been proved exactly in theory, but some reasons could be given. For example, one can see some calculations [4]. The triplet pairing is not discussed in this paper because it necessarily belong to the extent of low temperature superconductivity [5].

2. What is pseudogap?

From the difference between the metal, insulator and semiconductor we understand the bandgap; from superconductivities we remember the pairing gap. In the present, someone call other energy gaps as pseudogaps which usually start to appear above the superconducting transition temperature. Some authors have suggested that these gaps should be classified as the precursor pairing gap, the spin gap, the charge gap and so on. It seems that the precursor pairing gap is just the pseudogap. However, if there is indeed precursor pairing, the precursor pairing gap should locate far away from the Fermi level as suggested above, otherwise, if the pairing gap is around the Fermi level, it should be the superconducting gap. So far, one should find the differences between my ideas and the popular ideas of many physicists. One of the popular ideas is that the precursor pairing occurs when the pairs are not coherent or correlated. Because the gaps far away from the Fermi level do not affect the superconductivity, we suggest that they should be excluded from the pseudogaps. The pseudogaps should be the energy gaps which could appear above the superconducting transition temperature and locate around the Fermi level. We will focus on the pseudogap associated with the spin density wave (SDW) [6] in this paper. The pseudogap may be incomplete gap which could be due to spin fluctuations.

3. What is the relation between superconductivity and pseudogap?

If the pseudogap is associated with the SDW, it has been discussed in a previous paper [6]. When the spin correlation is strong enough, there is not the pairing gap but pseudogap, while there is not the pseudogap but pairing gap when the spin correlation is weak enough. In these cases, the superconductivity is independent of the pseudogap. As soon as the pairing gap coexists with the pseudogap, the competition between pairs and SDW occurs, and the pseudogap intends to decrease the superconducting transition temperature. However, both the pseudogap and high-Tc superconductivity require the strong correlation, the pseudogap looks like helping the high-Tc superconductivity, this is just why this question has been an open topic between physicists.

4. Is superconductivity related to magnetism?

The relation between superconductivity and magnetism has been a hot topic, and this can be observed in some articles [7, 8, 9, 10]. Magnetism and superconductivity are usually mutually exclusive, but one has felt that some secret may exist among the crossover region from magnetism to superconductivity. Particularly, one suggests that magnetic fluctuations may induce the high temperature superconductivity. Our work shows that the secret is that both magnetism and superconductivity favor the strongly correlated electron systems [6]. The electron correlation from strong to weak should be one of the causes of magnetic fluctuations, but the close relations between superconductivities and strong correlations are more easily found if we calculate the pairing gap in the wave vector space.

5. How to arrive at the highest temperature superconductivity?

When the pairing gap is calculated in the wave vector space, we find that the anisotropy of the electronic structure is a key to arrive at the high temperature superconductivity. One may have understood that the BCS theory is suitable for the common metals, and the metals are almost isotropic in physical properties. Here there is a secret that the isotropic materials necessarily are the low temperature superconductors according to our theory. If the metals are the weakly correlated ones again, they should be the phonon-mediated superconductors. Therefore, the BCS theory is suitable for the weakly correlated metals. Our works show that the electronic interactions are important for the strongly correlated superconductors. When the electron number is determined, to arrive at the highest temperature superconductors, the appropriate strong electron correlation and the evident anisotropy are the two key factors. The electron number, the appropriate strong correlation and the evident anisotropy are the decisive factors of superconductivity, and we will call it "three factor theory".

6. Is the qusi-electron number a conservation one?

Almost all of the physicists believe that the qusi -electron number is equal to the electron number, and the chemical potential of electron systems has been usually determined with this common sense. Particularly, the investigations of superconductivities have followed this sense. However, we indeed find the qusi-electron number < the electron number in a superconducting state [11]. Thus we suggest a fermion -boson model which keeps the electron number conservation. That is, the qusi-electron number + two times of boson number = the electron number. Of course, because the bosons correspond to the pairs around the Fermi level, thus two times of boson number << the electron number, and we have the equation: the qusi-electron number ≈ the electron number. The electron number is defined as the electron number of incompletely filled band(s). Moreover, because the basic features (zero resistance effect, Meissner effect, Josephson effect and so on) of superconductivity can be explained by the phenomenological theories and various microscopic theories of superconductivity necessarily take some approximation, the features of bosons of pairs have been ignored in the past. By the way, the electron number may not be equal to the carrier number.

7. Why do various superconductors have different Tc ?

Now let us have a look about the application of "three factor theory". Insulators are usually not superconductors because the electron number is equal to zero. The heavy fermion materials are the low temperature superconductors because their electronic properties are almost isotropic. The organic materials should not have high Tc, too. The good metals Au, Al and Cu are not superconductors because their electronic properties are almost isotropic and the electron-phonon interactions may be very weak (good metals certainly are weak correlated materials.), they should have very low superconducting transition temperatures, Tc → oK. However, because they are not single crystals, thus the structure disorders break their possible superconducting states. Now let us have a look about the Fermi surface. The energy bands of good metals are usually taken as the parabolic type, $E(k) \sim k^2$, their Fermi surfaces (Fermi level positions in the wave vector space) are like sphere shape. There should be the corresponding relation between the Fermi surface and the electronic property. The Fermi surface is sphere shape if the electronic properties are isotropic, and the Fermi surface deviates from the sphere shape when the electronic properties are anisotropic. The Fermi surfaces of 1D electron systems seriously deviates from the sphere shape, it seems that 1D materials could have the highest superconducting transition temperature. However, superconductivity is a macroscopic property, a 1D macroscopic material has not been found to my knowledge. The Fermi surfaces of 3D electron systems are not too far from the sphere shape, 2D materials with rectangular unit cells are the most possible election material, thus the cuprate materials (with CuO2 planes) are the high-Tc superconductors which intend to meet the "three factors theory". MgB2 [12] is not 2D material, the conductive layers of some iron pnictides [13] do not consisting of rectangular unit cells (each layer may not be in a plane), and some iron pnictides are not strongly correlated materials, thus their Tc are not high enough.

8. Are there relationships between superconductivities and other properties?

As discussed above, the approximate isotropic property corresponds to the very low Tc. The approximate isotropic property also corresponds to the s-wave pairing gap. Simply speaking, isotropic → low Tc → s wave symmetry. The pseudogap may appear in strongly correlated systems, and this means that the pseudogap could replace the superconducting gap. This gives the relation: isotropic → low T* → s wave symmetry. Some answers could be seen in answers 7 and 9.

9. What interaction is important for superconducting pairs?

The electron-phonon interactions are important for the weakly correlated materials while the electronic interactions are important for the strongly correlated materials.

The problem will be complex for the strongly correlated and approximately isotropic materials in which there may be the competition between superconducting states and SDW orders. Moreover, both the electronic interactions and the electron-phonons interactions may take part in the electron pairing. The electronic interactions refer to the effective electron-electron interactions which are affected by the interactions between electrons and other factors. "Other factors" include impurities, nucleu spins or ion spins, external fields, and so on. Because of these causes, the iron pnictides have complex isotope effects. The electronic interactions are usually repulsive, and one may questions how a repulsive interaction pull two electrons to a pair. Firstly, a pair is not a bound electron pair in the real space. The pair-breaking and the pair-forming continue to occur in the real space. Secondly, the superconducting pairs are around the Fermi surface, and the pair number << the (total) electron number. Thirdly, the repulsive interactions could be equivalent to the attractive interactions (for the pairs around the Fermi surface) in the energy gap equations. Some details will not be discussed in this paper.

10. What is the relation between isotope effect and interaction?

The isotopic effects of superconductivity are easy to understand when the electron-phonon interactions play an important part in a superconductor which can be described with the BCS theory. If the superconductivity is dominated by the electronic interactions, the isotope effects usually are not obvious, as shown in optimally doped cuprate superconductors. However, some underdoped cuprates have complex isotope effects [14,15,16], why? The isotope substitution can change the total spin of atomic nucleus, when the interactions between the electron spins and the nucleu spins (or ion spins) are considered, the energy band will be affected by the isotope substitution, thus superconductivity is affected. In addition, with the change of material structures (especially, with the changes among anisotropic and isotropic ones), the roles of electronic interactions may be decreased, while the roles of phonons may be increased, these will lead the isotope effect more complex.

In summary, to understand the high temperature superconductivity, it is necessary to correct our ideas.

1. Bardeen, J. Cooper, L. Schrieffer, J. R. Theory of superconductivity. Phys. Rev. 108, 1175-1204